\documentclass[a4paper,11pt]{article}
\usepackage{jheppub} % for details on the use of the package, please
\usepackage{subfigure}

\preprint{KUNS-2989}

\title{\boldmath 
The double scaling limit of randomly coupled Pauli XY spins}

\author{Ryota Watanabe}

\affiliation{Department of Physics, Kyoto University, Kyoto 606-8502, Japan}

\emailAdd{watanabe@gauge.scphys.kyoto-u.ac.jp}

\abstract{
We consider the double scaling limit of a model of Pauli spin operators recently studied in Hanada et al. \cite{Hanada:2023rkf} and evaluate the moments of the Hamiltonian by the chord diagrams.
We find that they coincide with those of the double scaled SYK model, which makes it more likely that this model may play an important role in the study of holography.
We compare the model with another previously studied model.
We also speculate on the form of the Hamiltonian in the double scaling limit.
}

\begin{document}

\maketitle
\flushbottom

\section{Introduction}
\label{sec:1}
In the study of holography \cite{Maldacena:1997re}, simple quantum mechanical models with gravity dual have been looked for.
A major breakthrough was made by the Sachdev-Ye-Kitaev (SYK) model \cite{Sachdev:1992fk,Kitaev:seminar}, which is a quantum mechanical model of $N_{\rm Maj}$ Majorana fermions $\psi_i$,
\begin{equation}
	H_{\rm SYK} = i^{p/2}\sum_{1\leq i_1<\cdots<i_p\leq N_{\rm Maj}} J_{i_1\cdots i_p}\psi_{i_1}\cdots\psi_{i_p}\,,
\label{eq:SYKp}
\end{equation}
where $\{\psi_i,\psi_j\}=2\delta_{ij}$ and the couplings $J_{i_1\cdots i_p}$ are independent random variables with
\begin{equation}
	\langle J_{i_1\cdots i_p} \rangle = 0\,,\quad \langle J_{i_1\cdots i_p}^2\rangle = \begin{pmatrix}N_{\rm Maj} \\ p\end{pmatrix}^{-1}\mathcal{J}^2\,.
\label{eq:SYKp coupling}
\end{equation}
$p$ is assumed to be an even integer.
Remarkably, the SYK model is solvable under the large-$N_{\rm Maj}$ limit and the disorder averages about the couplings \cite{Maldacena:2016hyu}. In addition, this model is known to saturate the bound on chaos \cite{Maldacena:2015waa}, and is expected to have a gravity dual.
Over the past decade, the SYK model has been investigated extensively and has deepen our understanding of holography. 
One of the interesting directions is the study of the model in the double scaling regime
\begin{equation}
	N_{\rm Maj}\to\infty\,, \quad p\to\infty\,, \quad \lambda = \frac{2p^2}{N_{\rm Maj}} = \text{fixed}\,,
\label{eq:SYKp DSL}
\end{equation}
where the SYK model can be solved analytically \cite{Cotler:2016fpe}.
In \cite{Berkooz:2018jqr}, the method of chord diagrams developed in \cite{Berkooz:2018qkz} was used to find the correlation functions in the double scaled SYK model (see also \cite{Okuyama:2022szh,Okuyama:2023bch,Okuyama:2023iwu,Okuyama:2023byh,Okuyama:2023kdo,Okuyama:2023aup,Okuyama:2023yat,Mukhametzhanov:2023tcg}).
Recently, the Krylov complexity \cite{Parker:2018yvk,Balasubramanian:2022tpr} was also studied in the double scaled SYK model \cite{Bhattacharjee:2022ave,Rabinovici:2023yex}.
Also, the relation between the double scaled SYK model and de Sitter spacetime is being investigated \cite{Susskind:2021esx,Susskind:2022dfz,Susskind:2022bia,Goel:2023svz,Narovlansky:2023lfz,Rahman:2023pgt}.

In order to better understand the SYK model and holography, it will be useful to consider generalizations and analogues of the model.
Various models have been studied such as the complex SYK model \cite{Sachdev:2015efa,Gu:2019jub,Berkooz:2020uly}, the supersymmetric SYK model \cite{Fu:2016vas,Li:2017hdt,Berkooz:2020xne,Biggs:2023mfn}, the higher dimensional generalizations \cite{Gu:2016oyy,Berkooz:2016cvq,Jian:2017unn,Jian:2017jfl,Song:2017pfw,Haldar:2017pyx}, generalizations on the way of coupling \cite{Xu:2020shn,Garcia-Garcia:2020cdo,Tezuka:2022mrr,Anegawa:2023vxq,Anegawa:2023tgp}, and the Pauli random spin models \cite{erdHos2014phase,Berkooz:2018qkz,Baldwin:2019dki,Hanada:2023rkf,Swingle:2023nvv}.
For example, in \cite{Hanada:2023rkf}, the authors considered a system of $N_{\rm spin}=N_{\rm Maj}/2$ spins with the following Hamiltonian:
\begin{equation}
	H_4 = \sqrt{\frac{6}{N_{\rm Maj}^3}}\sum_{1\leq a<b<c<d\leq N_{\rm Maj}} J_{abcd}\,i^{\eta_{abcd}}O_aO_bO_cO_d\,,
\label{eq:SpinXY4}
\end{equation}
where $O_a$'s are given by
\begin{align}
	O_1 &= \sigma_{1,x} = \sigma_x \otimes I \otimes I \otimes \cdots \otimes I \otimes I \notag \\
	O_2 &= \sigma_{1,y} = \sigma_y \otimes I \otimes I \otimes \cdots \otimes I \otimes I \notag \\
	O_3 &= \sigma_{2,x} = I \otimes \sigma_x \otimes I \otimes \cdots \otimes I \otimes I \notag \\
	O_4 &= \sigma_{2,y} = I \otimes \sigma_y \otimes I \otimes \cdots \otimes I \otimes I \notag \\
	\vdots & \notag \\
	O_{N_{\rm Maj}-1} &= \sigma_{N_{\rm spin},x} = I \otimes I \otimes I \otimes \cdots \otimes I \otimes \sigma_x \notag \\
	O_{N_{\rm Maj}} &= \sigma_{N_{\rm spin},y} = I \otimes I \otimes I \otimes \cdots \otimes I \otimes \sigma_y\,. 
\label{eq:SpinXY4 operator}
\end{align}
The coupling $J_{abcd}$ is a random variable sampled from
\begin{equation}
	P(J_{abcd}) = \frac{1}{\sqrt{2\pi}}e^{-J_{abcd}^2/2}\,,
\label{eq:J distribution}
\end{equation}
and $\eta_{abcd}$ is the number of spins in a given $(a,b,c,d)$ in which both $x$ and $y$ components appear.
Although this model resembles those in \cite{erdHos2014phase,Berkooz:2018qkz,Baldwin:2019dki,Swingle:2023nvv}, there are some differences.
(1) Only $\sigma_x$ and $\sigma_y$ appear in \eqref{eq:SpinXY4}.
(2) Both $\sigma_x,\,\sigma_y$ can act on the same spin simultaneously in one interaction term.
For the latter, $\sigma_x\sigma_y$ could be replaced by $i\sigma_z$. 
Then all of $\sigma_x, \sigma_y, \sigma_z$ would appear, but the interaction term including $\sigma_z$ would be shorter in length than the other terms and still slightly different from the spin model considered so far.
The model \eqref{eq:SpinXY4} was studied numerically in \cite{Hanada:2023rkf}, and results very close to those of the usual SYK model were confirmed for the density of states, level statistics, spectral form factors, and two-point functions.

In this paper, we study a generalization of the above model.
The Hamiltonian we consider is given by
\begin{equation}
	H_p = \mathcal{N}\sum_{1\leq a_1<\cdots<a_p\leq N_{\rm Maj}} J_{a_1\cdots a_p}\,i^{\eta_{a_1\cdots a_p}}O_{a_1}\cdots O_{a_p}\,, \quad \mathcal{N}=\begin{pmatrix}N_{\rm Maj} \\ p\end{pmatrix}^{-1/2}\,,
\label{eq:SpinXYp}
\end{equation}
which is called SpinXYp in \cite{Hanada:2023rkf}. Here, $O_a$'s are given in the same way as \eqref{eq:SpinXY4 operator}.
The coupling $J_{a_1\cdots a_p}$ is a random variable sampled from
\begin{equation}
	P(J_{a_1\cdots a_p}) = \frac{1}{\sqrt{2\pi}}e^{-J_{a_1\cdots a_p}^2/2}\,,
\label{eq:J distribution in SpinXYp}
\end{equation}
and $\eta_{a_1\cdots a_p}$ is the number of spins in a given $(a_1,\cdots,a_p)$ in which both $x$ and $y$ components appear.
We study this model in the double scaling limit using the chord diagram method of \cite{Berkooz:2018qkz,Berkooz:2018jqr,Berkooz:2020xne,Berkooz:2020uly}.
The results confirm that the disorder average of the moments of the Hamiltonian is in perfect agreement with the results in the double scaling limit of the usual SYK model.
This is as expected from the numerical results of \cite{Hanada:2023rkf}, and suggests that the SpinXYp model can play as important a role as the SYK model in the study of holography.

This paper is organized as follows.
In Sec.~\ref{sec:2}, we review the chord diagram method in the SYK model based on \cite{Berkooz:2018jqr}.
In Sec.~\ref{sec:3}, we consider the double scaling limit in the SpinXYp model \eqref{eq:SpinXYp}.
Based on the chord diagram method reviewed in Sec.~\ref{sec:2}, we evaluate the chord diagrams of the SpinXYp model in the double scaling limit in Sec.~\ref{sec:3-1}.
In Sec.~\ref{sec:3-2}, we compare the result obtained in Sec.~\ref{sec:3-1} with that of another Pauli spin model of \cite{Berkooz:2018qkz}.
As pointed out above, one feature of the SpinXY model is that both $\sigma_x$ and $\sigma_y$ can act on the same spin simultaneously in a single interaction term.
We are interested in how often such an event occurs in the double scaling limit.
Therefore, we calculate in Sec.~\ref{sec:3-3} the expectation value of $\eta_{a_1\cdots a_p}$ for each term of \eqref{eq:SpinXYp} in the double scaling limit.
Sec.~\ref{sec:4} is devoted to a summary and discussions.

%%%%%%%%%%%%%%%%%%%%%%%%%%%%%%%%%%%%%%%%%%%%%%%%%%%

\section{Review of the chord diagram method}
\label{sec:2}
In this section, we review the calculation of the moments of the SYK Hamiltonian \eqref{eq:SYKp}
\begin{equation}
	m_n^{({\rm SYK})} \equiv \langle{\rm Tr} H_{\rm SYK}^n\rangle
\label{eq:moment 0}
\end{equation}
by the chord diagram method based on \cite{Berkooz:2018jqr}.
We set ${\rm Tr}[I]=1$ and $\mathcal{J}=1$ in the following.

Let us denote the indices $(i_1,\cdots,i_p)$ in ascending order by $I$, and write the Hamiltonian \eqref{eq:SYKp} as
\begin{equation}
	H_{\rm SYK} = i^{p/2}\sum_{I}J_I\psi_I\,, \quad \psi_I\equiv \psi_{i_1}\psi_{i_2}\cdots \psi_{i_p}\,.
\label{eq:shorthand Hamiltonian}
\end{equation}
Then, the moment \eqref{eq:moment 0} becomes
\begin{equation}
	m_n^{({\rm SYK})} = i^{pn/2}\sum_{I_1,\cdots,I_n}\langle J_{I_1}\cdots J_{I_n} \rangle {\rm Tr}[\psi_{I_1}\cdots \psi_{I_n}]\,.
\label{eq:moment 1}
\end{equation}
Since $J_I$ follows an independent Gaussian distribution with zero mean, we obtain
\begin{equation}
	\langle J_{I_1}\cdots J_{I_n} \rangle = 0 \quad (n=\text{odd})\,.
\end{equation}
Therefore, in the following, we consider only the case $n=2k~(k=1,2,\cdots)$.
Also, for convenience, we scale the coupling $J_I$ as
\begin{equation}
	J_I \to \mathcal{N}J_{I}\,, \quad \mathcal{N} = \begin{pmatrix}N_{\rm Maj} \\ p\end{pmatrix}^{-1/2}
\end{equation}
and rewrite \eqref{eq:moment 1} as
\begin{equation}
	m_{2k}^{({\rm SYK})} = i^{pk}\mathcal{N}^{2k}\sum_{I_1,\cdots,I_{2k}}\langle J_{I_1}\cdots J_{I_{2k}} \rangle {\rm Tr}[\psi_{I_1}\cdots \psi_{I_{2k}}]\,,
\label{eq:moment 2}
\end{equation}
where $\langle J_I\rangle = 0$ and $\langle J_I^2\rangle = 1$, especially independent of $N_{\rm Maj}$. 

Next, we consider which terms in \eqref{eq:moment 2} survive in the limit $N_{\rm Maj}\to\infty$.
First, if there is an index in $I_1,\cdots,I_{2k}$ that is different from all the rest, then $\langle J_{I_1}\cdots J_{I_{2k}} \rangle = 0$ since $\langle J_I\rangle = 0$.
Therefore, the indices in  $I_1,\cdots,I_{2k}$ must form pairs or groups.
It follows that if $m$ is the number of independent elements in $(I_1,\cdots,I_{2k})$, then $m\leq k$ for the surviving terms.
Now, we rewrite \eqref{eq:moment 2} as follows:
\begin{equation}
	m_{2k}^{({\rm SYK})} = i^{pk}\mathcal{N}^{2k}\sum_{m=1}^k \sum_{\text{partitions}}\sum_{\{I\}}~\langle J_{I_1}\cdots J_{I_{2k}} \rangle {\rm Tr}[\psi_{I_1}\cdots \psi_{I_{2k}}]\,,
\label{eq:moment 3}
\end{equation}
where $\sum_{\text{partitions}}$ is the sum over all ways to devide $\{1,2,\cdots,2k\}$ into $m$ groups, and $\sum_{\{I\}}$ is the sum over $I_1,\cdots,I_{2k}$ respecting the groupings.
For example, if $k=2$ and $\{1,2,3,4\} \to\{\{1,2\},\{3,4\}\}$ is considered as partition, then $\sum_{\{I\}}=\sum_{I_1,\cdots,I_4}\delta_{I_1I_2}\delta_{I_3I_4}$. 

\begin{figure}[t]
\centering
    \subfigure[]
     {\includegraphics[width=3cm]{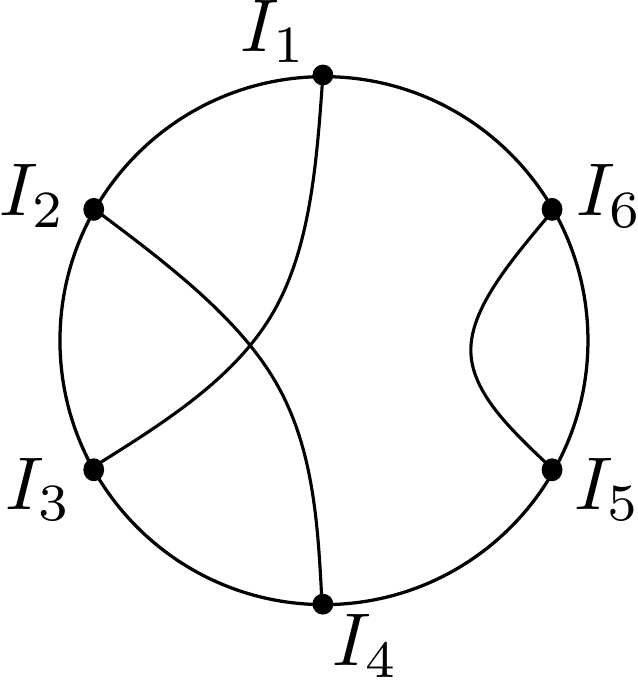} \label{fig:pairing1}}
    \hspace{20mm}
    \subfigure[]
     {\includegraphics[width=3cm]{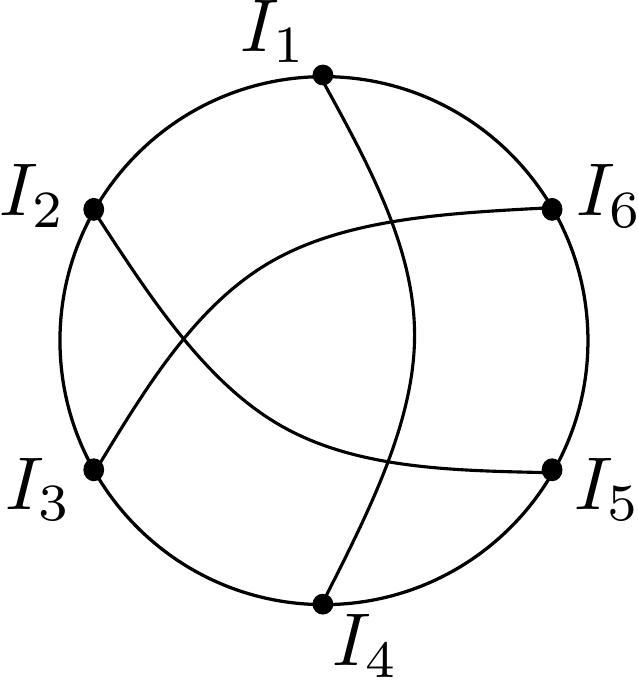} \label{fig:pairing2}}
    \caption{Examples of the chord diagrams in the case of $k=3$. (a) $I_1=I_3,\,I_2=I_4,\,I_5=I_6$. (b) $I_1=I_4,\,I_2=I_5,\,I_3=I_6$.}
\label{fig:chord diagram}
\end{figure}

Now, if we evaluate the order of the summation part of \eqref{eq:moment 3} in $N_{\rm Maj}\to\infty$, we can find that $\sum_{m=1}^k \sum_{\text{partitions}}$ does not depend on $N_{\rm Maj}$, but $\sum_{\{I\}}$ depends on $N_{\rm Maj}$ as
\begin{equation}
	\sum_{\{I\}} \sim \begin{pmatrix}N_{\rm Maj} \\ p\end{pmatrix}^m
\label{eq:order of I summation}
\end{equation}
in the leading order.
Therefore, the order of \eqref{eq:moment 3} in $N_{\rm Maj}$ becomes
\begin{equation}
	m_{2k}^{({\rm SYK})} \sim \mathcal{N}^{2k}\sum_{m=1}^k \begin{pmatrix}N_{\rm Maj} \\ p\end{pmatrix}^m = \sum_{m=1}^k \begin{pmatrix}N_{\rm Maj} \\ p\end{pmatrix}^{m-k}\,.
\label{eq:order of moment}
\end{equation}
If we take $N_{\rm Maj}\to\infty$, it can be seen that only the $m=k$ terms in $\sum_{m=1}^k$ survive, and the other $m<k$ terms become zero.
Eventually, we see that, in the limit $N_{\rm Maj}\to\infty$, only terms such that each of $I_1,\cdots,I_{2k}$ forms a pair can give a non-zero contribution in \ref{eq:moment 1}.
Then, we obtain
\begin{align}
	m_{2k}^{({\rm SYK})} &\sim i^{pk}\mathcal{N}^{2k}\sum_{\text{pairings}}\sum_{\{I\}}~\langle J_{I_1}\cdots J_{I_{2k}} \rangle {\rm Tr}[\psi_{I_1}\cdots \psi_{I_{2k}}] \notag \\
	&= i^{pk}\sum_{\text{pairings}}\mathcal{N}^{2k}\sum_{\{I\}}~ {\rm Tr}[\psi_{I_1}\cdots \psi_{I_{2k}}]\,.
\label{eq:pairings}
\end{align}
It is useful to represent the each term in \eqref{eq:pairings} as the chord diagram in Fig.~\ref{fig:chord diagram}. 
The circle in the figure represents ${\rm Tr}$, and the points represent each $\psi_I$.
The line (chord) connecting the dots indicates that the endpoints have a common index.

Let us consider Fig.~\ref{fig:pairing1}.
In this diagram, $I_5$ and $I_6$ are adjacent with a common index, and we can evaluate their contribution as
\begin{equation}
	\psi_{I_5}\psi_{I_6}\big|_{I_5=I_6} = i^p \quad (p=\text{even})\,.
\end{equation}
Note that the sum for $I_5\,(=I_6)$ in $\sum_{\{I\}}$ is trivial and cancels out with the overall factor $\mathcal{N}^2$ (in the double scaling limit).
To evaluate the rest of Fig.~\ref{fig:pairing1}, we need to exchange $\psi_I$ appropriately so that points with a common index become adjacent.
Let us consider swapping $\psi_{I_1}$ and $\psi_{I_2}$.
If we denote by $r(I_1,I_2)$ the number of common $\psi_i$'s contained in $\psi_{I_1}$ and $\psi_{I_2}$,
\begin{equation}
	\psi_{I_1}\psi_{I_2} = (-1)^{r(I_1,I_2)}\psi_{I_2}\psi_{I_1}\,.
\end{equation}
Then, Fig.~\ref{fig:pairing1} in the double scaling limit becomes
\begin{equation}
	\mathcal{N}^2\sum_{I_1}\mathcal{N}^2\sum_{I_2}(-1)^{r(I_1,I_2)}\,.
\end{equation}
Since $r(I_1,I_2)$ follows a Poisson distribution
\begin{equation}
	P(r) = \frac{1}{r!}\left(\frac{p^2}{N_{\rm Maj}}\right)^r e^{-p^2/N_{\rm Maj}}
\label{eq:Poisson}
\end{equation}
under double scaling limit,\footnote{See the lemma 9 in \cite{erdHos2014phase} and Appendix A of \cite{Berkooz:2018jqr}.} $\mathcal{N}^2\sum_{I_2}(-1)^{r(I_1,I_2)}$ can be evaluated as follows:
\begin{equation}
	\mathcal{N}^2\sum_{I_2}(-1)^{r(I_1,I_2)} \sim \sum_{r=0}^\infty\frac{1}{r!}\left(\frac{p^2}{N_{\rm Maj}}\right)^r e^{-p^2/N_{\rm Maj}}(-1)^r = e^{-\lambda}\equiv q\,.
\end{equation}
The remaining $\mathcal{N}^2\sum_{I_1}$ is simply 1, and finally the contribution of Fig.~\ref{fig:pairing1} becomes $q$.

Next, we consider Fig.~\ref{fig:pairing2}.
Similarly to Fig.~\ref{fig:pairing1}, if we swap the $\psi_I$'s as $I_1\leftrightarrow I_2,\,I_3\leftrightarrow I_4\,(=I_1),\,I_5\,(=I_2)\leftrightarrow I_6\,(=I_3)$, we obtain
\begin{equation}
	\mathcal{N}^6\sum_{I_1,I_2,I_3}(-1)^{r(I_1,I_2)}(-1)^{r(I_2,I_3)}(-1)^{r(I_3,I_1)}\,.
\end{equation}
Now let us consider first fixing $I_1, I_2$ and adding up for $I_3$:
\begin{equation}
	\mathcal{N}^4\sum_{I_1,I_2}(-1)^{r(I_1,I_2)}\mathcal{N}^2\sum_{I_3}(-1)^{r(I_2,I_3)}(-1)^{r(I_3,I_1)}\,.
\end{equation}
As mentioned in \cite{erdHos2014phase,Berkooz:2018qkz}, the probability that $\psi_{I_1}, \psi_{I_2}$ and $\psi_{I_3}$ share a common $\psi_i$ simultaneously becomes zero in the double scaling limit.
Therefore, we can treat $r(I_2,I_3)$ and $r(I_3,I_1)$ as independent random variables.
Then, we find
\begin{equation}
	\mathcal{N}^2\sum_{I_3}(-1)^{r(I_2,I_3)}(-1)^{r(I_3,I_1)} \sim q^2\,.
\end{equation}
Evaluating the rest, we can evaluate Fig.~\ref{fig:pairing2} as $q^3$.

Repeating the above procedures, we see that in general the moment $m_{2k}$ can be expressed in the double scaling limit as
\begin{equation}
	m_{2k}^{({\rm SYK})} = \sum_{\text{chord diagrams}}q^{\#~\text{intersections}}\,.
\label{eq:result of SYK}
\end{equation}

%%%%%%%%%%%%%%%%%%%%%%%%%%%%%%%%%%%%%%%%%%%%%%%%%%%%%%%%%%%

\section{The double scaling limit of the SpinXYp model}
\label{sec:3}
In this section, we study the double scaling limit
\begin{equation}
	N_{\rm Maj}\to\infty\,, \quad p\to\infty\,, \quad \lambda = \frac{2p^2}{N_{\rm Maj}} = \text{fixed}\,,
\label{eq:SpinXYp DSL}
\end{equation}
of the SpinXYp model \eqref{eq:SpinXYp}.
Although this model resembles the spin models considered in \cite{erdHos2014phase,Berkooz:2018qkz,Baldwin:2019dki,Swingle:2023nvv}, it differs in some points:
\begin{itemize}
\item Only $\sigma_x$ and $\sigma_y$ appear in \eqref{eq:SpinXYp}.
\item Both $\sigma_x,\,\sigma_y$ can act on the same spin simultaneously in one interaction term.
\end{itemize}
We shall refer to the latter event as collision in the following.

%%%%%%%%%%%%%%%%%%%%%%%%%%%%%%%%%%%%%%%%%%%%%%%%%%%%%%%%%%%

\subsection{Evaluation of the moment of the Hamiltonian}
\label{sec:3-1}
The moment of the Hamiltonian \eqref{eq:SpinXYp}
\begin{equation}
	m_n^{(p)} = \langle{\rm Tr} H_p^n\rangle
\label{eq:SpinXYp moment 0}
\end{equation}
can be evaluated by the method in Sec.~\ref{sec:2}.
We will see below that the calculation process is not affected by the collisions at all.

To begin with, let us denote the indices $(a_1,\cdots,a_p)$ in ascending order by $I$, and write the Hamiltonian \eqref{eq:SpinXYp} as
\begin{equation}
	H_p = \mathcal{N}\sum_{I}J_I\,i^{\eta_I}O_I\,, \quad O_I\equiv O_{a_1}O_{a_2}\cdots O_{a_p}\,.
\end{equation}
Substituting this into \eqref{eq:SpinXYp moment 0}, we find by the same argument as Sec.~\ref{sec:2} that in the double scaling limit
\begin{equation}
	m_{2k}^{(p)} \sim \sum_{\text{pairings}}\mathcal{N}^{2k}\sum_{\{I\}}~ {\rm Tr}[i^{\eta_{I_1}}O_{I_1}\cdots i^{\eta_{I_{2k}}}O_{I_{2k}}]\,.
\label{eq:SpinXYp pairings}
\end{equation}
As in Sec.~\ref{sec:2}, each term of $\sum_{\text{pairings}}$ is represented as in Fig.~\ref{fig:chord diagram}.
Note that the points on the circle represent each $i^{\eta_{I}}O_{I}$ in the present case.

Now, let us consider Fig.~\ref{fig:pairing1} in the case of the SpinXYp model.
In this diagram, $I_5$ and $I_6$ are adjacent with a common index, and we can evaluate them as
\begin{equation}
	i^{\eta_{I_5}}O_{I_5}i^{\eta_{I_6}}O_{I_6}\big|_{I_5=I_6} = 1
\end{equation}
Again, note that the sum for $I_5\,(=I_6)$ in $\sum_{\{I\}}$ is trivial and cancels out with the overall factor $\mathcal{N}^2$ (in the double scaling limit).
To evaluate the rest of Fig.~\ref{fig:pairing1}, we need to exchange $O_I$ appropriately so that points with a common index become adjacent.
Let us consider swapping $O_{I_1}$ and $O_{I_2}$.
If $O_{I_1}$ contains $r(I_1,I_2)$ Pauli matrices that anti-commute with the Pauli matrices that make up $O_{I_2}$, we have
\begin{equation}
	O_{I_1}O_{I_2} = (-1)^{r(I_1,I_2)}O_{I_2}O_{I_1}\,.
\end{equation}
Then, the contribution of Fig.~\ref{fig:pairing1} in the double scaling limit becomes
\begin{equation}
	\mathcal{N}^2\sum_{I_1}\mathcal{N}^2\sum_{I_2}(-1)^{r(I_1,I_2)}\,.
\label{eq:contribution 1}
\end{equation}
Since we can see that $r(I_1,I_2)$ follows the same Poisson distribution as \eqref{eq:Poisson}
under double scaling limit,\footnote{For a given Pauli matrix, the Pauli matrix that anti-commutes with it is uniquely determined in the present case. Therefore, the same argument as Appendix A of \cite{Berkooz:2018jqr} holds.}
we again have
\begin{equation}
	\mathcal{N}^2\sum_{I_2}(-1)^{r(I_1,I_2)} \sim \sum_{r=0}^\infty\frac{1}{r!}\left(\frac{p^2}{N_{\rm Maj}}\right)^r e^{-p^2/N_{\rm Maj}}(-1)^r = e^{-\lambda}\equiv q\,.
\end{equation}
The remaining $\mathcal{N}^2\sum_{I_1}$ is simply 1, and \eqref{eq:contribution 1} becomes $q$.

Next, we consider Fig.~\ref{fig:pairing2}.
In the same way as in Sec.~\ref{sec:2}, if we swap the $O_I$'s as $I_1\leftrightarrow I_2,\,I_3\leftrightarrow I_4\,(=I_1),\,I_5\,(=I_2)\leftrightarrow I_6\,(=I_3)$, we have
\begin{equation}
	\mathcal{N}^6\sum_{I_1,I_2,I_3}(-1)^{r(I_1,I_2)}(-1)^{r(I_2,I_3)}(-1)^{r(I_3,I_1)}\,.
\end{equation}
Then, let us consider first fixing $I_1, I_2$ and sum over $I_3$:
\begin{equation}
	\mathcal{N}^4\sum_{I_1,I_2}(-1)^{r(I_1,I_2)}\mathcal{N}^2\sum_{I_3}(-1)^{r(I_2,I_3)}(-1)^{r(I_3,I_1)}\,.
\label{eq:contribution 2}
\end{equation}
Similarly to \cite{erdHos2014phase,Berkooz:2018qkz}, the probability that $O_{I_3}$ anti-commutes with $O_{I_1}$ and $O_{I_2}$ at exactly the same Pauli matrices becomes zero in the double scaling limit.\footnote{The logic is the same as in \cite{Berkooz:2018qkz}. The expectation value of the number of common elements of $I_1$ and $I_2$ is equal to $p^2/N_{\rm Maj}$. The expectation value of the number of elements in $I_3$ that anti-commutes with those $p^2/N_{\rm Maj}$ elements is $p^2/N_{\rm Maj}\times p/N_{\rm Maj}$, which is zero in the double scaling limit.}
Therefore, we may treat $r(I_2,I_3)$ and $r(I_3,I_1)$ as independent random variables.
Then, in the same way as Sec.~\ref{sec:2}, \eqref{eq:contribution 2} becomes $q^3$.

As in Sec.~\ref{sec:2}, we see that in general the moment \eqref{eq:SpinXYp moment 0} can be expressed in the double scaling limit as
\begin{equation}
	m_{2k}^{(p)} = \sum_{\text{chord diagrams}}q^{\#~\text{intersections}}\,.
\label{eq:SpinXYp moments result}
\end{equation}
This is exactly the same as the result \eqref{eq:result of SYK} of the double scaled SYK model.

\subsection{Comparison with another model}
\label{sec:3-2}
Now, we would like to make a comparison with the another model of Pauli spins \cite{Berkooz:2018qkz}.
First, we briefly review their evaluation of the chord diagram.
The Hamiltonian is given by
\begin{equation}
	\tilde{H}=3^{-p/2}\begin{pmatrix} N_{\rm spin} \\ p \end{pmatrix}^{-1/2} \sum_{1\leq i_1<\cdots<i_p\leq N_{\rm spin}}\sum_{a_1,\cdots,a_p=1}^3 \alpha_{a_1,\cdots,a_p,(i_1,\cdots,i_p)}\sigma_{i_1,a_1}\cdots \sigma_{i_p,a_p}\,,
\end{equation}
where $\sigma_{i,a}$ is the Pauli matrix $\sigma_a~(a=1,2,3)$ acting on the $i$-th spin, and $\alpha$ are independent random coupling with zero mean and unit standard deviation.
All of $\sigma_x, \sigma_y, \sigma_z$ appear in this model, and at most one Pauli matrix acts on one spin in one interaction term.
As shown in \cite{Berkooz:2018qkz}, we can evaluate the moments of the Hamiltonian via chord diagrams as
\begin{equation}
	\langle {\rm Tr}\tilde{H}^{2k} \rangle = \sum_{\text{chord diagrams}}\tilde{q}^{\#~\text{intersections}}\,, \quad \tilde{q} = e^{-\tilde{\lambda}}\,, \quad \tilde{\lambda} \equiv \frac{4}{3}\frac{p^2}{N_{\rm spin}}
\label{eq:Berkooz moments}
\end{equation}
in the double scaling limit.
The derivation is as follows.
Suppose that two chords intersect in a given chord diagram.
Then, they give a factor
\begin{equation}
	3^{-2p}\begin{pmatrix} N_{\rm spin} \\ p \end{pmatrix}^{-2}\sum_{I,J}{\rm Tr}(\sigma_I\sigma_J\sigma_I\sigma_J)\,,
\end{equation}
where $\sigma_I$ is a shorthand notation of $\sigma_{i_1,a_1}\cdots \sigma_{i_p,a_p}$.
If there are overlaps of spins in $I$ and $J$, then $\sigma_I$ and $\sigma_J$ do not necessarily commute.
Each overlap gives a factor
\begin{equation}
	3^{-2}\sum_{a,b=1}^3{\rm Tr}(\sigma_a\sigma_b\sigma_a\sigma_b)=-\frac{1}{3}\,.
\label{eq:spin overlap}
\end{equation}
Since the number $r$ of overlap of spins in the double scaling limit follows a Poisson distribution
\begin{equation}
	P(r) = \frac{1}{r!}\left(\frac{p^2}{N_{\rm spin}}\right)^r e^{-p^2/N_{\rm spin}}\,,
\label{eq:spin Poisson}
\end{equation}
each intersection of two chords gives
\begin{equation}
	\sum_{r=0}^\infty P(r) \left(-\frac{1}{3}\right)^r = e^{-\tilde{\lambda}}\,.
\end{equation}
Therefore, the moments of the Hamiltonian can be evaluated as in \eqref{eq:Berkooz moments}.

Now, let us consider a similar model with only $\sigma_x$ and $\sigma_y$.
The Hamiltonian is given by
\begin{equation}
	H=2^{-p/2}\begin{pmatrix} N_{\rm spin} \\ p \end{pmatrix}^{-1/2} \sum_{1\leq i_1<\cdots<i_p\leq N_{\rm spin}}\sum_{a_1,\cdots,a_p=1}^2 \alpha_{a_1,\cdots,a_p,(i_1,\cdots,i_p)}\sigma_{i_1,a_1}\cdots \sigma_{i_p,a_p}\,.
\label{eq:new hamiltonian}
\end{equation}
In this case, an intersection of two chords in a chord diagram gives a factor
\begin{equation}
	2^{-2p}\begin{pmatrix} N_{\rm spin} \\ p \end{pmatrix}^{-2}\sum_{I,J}{\rm Tr}(\sigma_I\sigma_J\sigma_I\sigma_J)\,.
\end{equation}
Note that, instead of \eqref{eq:spin overlap}, each overlap gives zero
\begin{equation}
	2^{-2}\sum_{a,b=1}^2{\rm Tr}(\sigma_a\sigma_b\sigma_a\sigma_b)=0\,.
\end{equation}
Therefore, each intersection of two chords yields
\begin{equation}
	\sum_{r=0}^\infty P(r) \delta_{r0} = e^{-p^2/N_{\rm spin}}\,,
\end{equation}
and the moments of the Hamiltonian \eqref{eq:new hamiltonian} becomes
\begin{equation}
	\langle {\rm Tr}H^{2k} \rangle = \sum_{\text{chord diagrams}}\left(e^{-p^2/N_{\rm spin}}\right)^{\#~\text{intersections}}\,.
\end{equation}
Note that if we write $N_{\rm spin}$ as $N_{\rm Maj}/2$, this result is apparently identical to that of the SpinXYp model \eqref{eq:SpinXYp moments result}.

%%%%%%%%%%%%%%%%%%%%%%%%%%%%%%%%%%%%%%%%%%%%%%%%%%%%%%%%%%%

\subsection{The SpinXYp Hamiltonian in the double scaling limit}
\label{sec:3-3}
One of the feature of the SpinXYp model is that both $\sigma_x,\,\sigma_y$ can act on the same spin simultaneously in one interaction term, which we called collision.
The number of collisions in a given term in \eqref{eq:SpinXYp} is given by $\eta_{a_1\cdots a_p}$.
We can estimate the expectation value of the number of collisions in the double scaling limit.
The total number of events in which $\eta$ times collision occurs is given by
\begin{equation}
	\begin{pmatrix}N_{\rm spin} \\ \eta\end{pmatrix} \begin{pmatrix}N_{\rm spin}-\eta \\ p-2\eta\end{pmatrix} \times 2^{p-2\eta}\,.
\label{eq:collision events}
\end{equation}
The first factor $\begin{pmatrix}N_{\rm spin} \\ \eta\end{pmatrix}$ represents the total number of ways to choose the sites where the collisions occur.
The second factor $\begin{pmatrix}N_{\rm spin}-\eta \\ p-2\eta\end{pmatrix}$ corresponds to the total number of ways to select the sites where the operator acts without causing collision, and the last factor $2^{p-2\eta}$ is a factor of whether $\sigma_x$ or $\sigma_y$ acts on those sites.
Therefore, the probability distribution of $\eta$ is gven by
\begin{equation}
	P(\eta) = \frac{\begin{pmatrix}N_{\rm spin} \\ \eta\end{pmatrix} \begin{pmatrix}N_{\rm spin}-\eta \\ p-2\eta\end{pmatrix} \times 2^{p-2\eta}}{\begin{pmatrix}N_{\rm Maj} \\ p\end{pmatrix}}\,.
\label{eq:distribution of eta}
\end{equation}
In the double scaling limit, \eqref{eq:distribution of eta} becomes (see App.~\ref{app:1})
\begin{equation}
	P(\eta) \sim \frac{1}{\eta!}\left(\frac{p^2}{4N_{\rm spin}}\right)^\eta e^{-p^2/4N_{\rm spin}}\,.
\label{eq:distribution of eta in DSL}
\end{equation}
Therefore, the expectation value of $\eta$ in the double scaling limit is $p^2/4N_{\rm spin}$. 
Since this is a finite value, we naively expect that collision rarely occurs in the double scaling limit where the length $p$ of the interaction is brought to infinity.
Then, the system will be reduced to those of \cite{erdHos2014phase,Berkooz:2018qkz,Baldwin:2019dki,Swingle:2023nvv} with only $\sigma_x$ and $\sigma_y$.
This is consistent with the observation of Sec.~\ref{sec:3-2}.

%%%%%%%%%%%%%%%%%%%%%%%%%%%%%%%%%%%%%%%%%%%%%%%%%%%%%%%%%%%

\section{Summary}
\label{sec:4}
In this paper, we have calculated the moments of the SpinXYp Hamiltonian in the double scaling limit by the chord diagram method and found that they are in perfect agreement with those of the double scaled SYK model.
This model differs from those in \cite{erdHos2014phase,Berkooz:2018qkz,Baldwin:2019dki,Swingle:2023nvv} in the following points:
(1) only $\sigma_x$ and $\sigma_y$ appear in \eqref{eq:SpinXY4},
(2) both $\sigma_x,\,\sigma_y$ can act on the same spin simultaneously in one interaction term.
However, these differences do not prevent the calculation of the chord diagrams, and essentially the exact same method can be applied as in the case of the SYK model in \cite{Berkooz:2018jqr}.
Moreover, from the results of Sec.~\ref{sec:3-3}, it is naively expected that the Hamiltonian of the SpinXYp model approaches the Pauli spin model of the \cite{erdHos2014phase,Berkooz:2018qkz} with only $\sigma_x$ and $\sigma_y$ in the double scaling limit.

For the chord diagram to be easily evaluated, it is important that the square of each Hamiltonian interaction term is proportional to the identity and that the interaction terms commute with each other up to the sign factor.
In addition, the probability distribution of the sign factor must be known in a simple form.
If these conditions are not met, then an extra term will be generated each time the chord is interchanged, or the summation cannot be done analytically, and the evaluation of the trace would be difficult.
The SpinXYp model satisfies this requirement, and it is possible to evaluate the chord diagrams.
It is an interesting problem to evaluate the chord diagram with other models that satisfy this requirement.
The fact that the moments of the SpinXYp Hamiltonian in the double scaling limit coincide with that of the double scaled SYK model means, in particular, that the energy spectrum coincides.
This makes it increasingly likely that the SpinXYp model will play an important role in the study of holography.

%%%%%%%%%%%%%%%%%%%%%%%%%%%%%%%%%%%%%%%%%%%%%%%%%%%%%%%%%%%

\section*{Acknowledgments}
The author would like to thank Koji Hashimoto for valuable discussions.
The work of R.~W.~was supported by Grant-in-Aid for JSPS Fellows No.~JP22KJ1940.

%%%%%%%%%%%%%%%%%%%%%%%%%%%%%%%%%%%%%%%%%%%%%%%%%%%%%%%%%%%
\newpage

\appendix
\section{Derivation of \eqref{eq:distribution of eta in DSL}}
\label{app:1}
For simplicity, we write $N_{\rm spin}$ as $N$. 
Note that $N_{\rm Maj}=2N$, so \eqref{eq:distribution of eta} becomes
\begin{align*}
	P(\eta)
	&= \frac{\begin{pmatrix}N \\ \eta\end{pmatrix} \begin{pmatrix}N-\eta \\ p-2\eta\end{pmatrix} \times 2^{p-2\eta}}{\begin{pmatrix}2N \\ p\end{pmatrix}} \\
	&= \frac{N!}{\eta!(N-\eta)!}\frac{(N-\eta)!}{(p-2\eta)!(N-p+\eta)!}\frac{p!(2N-p)!}{(2N)!}2^{p-2\eta} \\
	&= \frac{1}{\eta!}\frac{p!}{(p-2\eta)!}\frac{(N-p)!}{(N-p+\eta)!}\frac{N!}{(N-p)!}\frac{(2N-p)!}{(2N)!}2^{p-2\eta} \\
	&= \frac{1}{\eta!}\frac{p!}{(p-2\eta)!}\frac{(N-p)!}{(N-p+\eta)!}\frac{\begin{pmatrix}N \\ p\end{pmatrix}}{\begin{pmatrix}2N \\ p\end{pmatrix}}2^{p-2\eta} \\
	&\sim \frac{1}{\eta!}\frac{p^{2\eta}}{N^\eta}\frac{\frac{N^p}{p!}e^{-p^2/2N}}{\frac{(2N)^p}{p!}e^{-p^2/4N}}2^{p-2\eta} \\
	&= \frac{1}{\eta!}\left(\frac{p^2}{4N}\right)^\eta e^{-p^2/4N}\,.
\end{align*}
Here, we used the asymptotic formula \cite{spencerasymptopia}
\begin{equation}
	\begin{pmatrix}N \\ p\end{pmatrix} \sim \frac{N^p}{p!}e^{-p^2/2N}\,,
\end{equation}
where $N\gg1,\,p=o(N^{2/3})$.

\bibliography{ref}

\begin{thebibliography}{10}

\bibitem{Hanada:2023rkf}
Masanori Hanada, Antal Jevicki, Xianlong Liu, Enrico Rinaldi, and Masaki
  Tezuka.
\newblock {A model of randomly-coupled Pauli spins}.
\newblock 9 2023.

\bibitem{Maldacena:1997re}
Juan~Martin Maldacena.
\newblock {The Large N limit of superconformal field theories and
  supergravity}.
\newblock {\em Adv. Theor. Math. Phys.}, Vol.~2, pp. 231--252, 1998.

\bibitem{Sachdev:1992fk}
Subir Sachdev and Jinwu Ye.
\newblock {Gapless spin fluid ground state in a random, quantum Heisenberg
  magnet}.
\newblock {\em Phys. Rev. Lett.}, Vol.~70, p. 3339, 1993.

\bibitem{Kitaev:seminar}
A.~Kitaev.
\newblock {A simple model of quantum holography}.
\newblock {\em talks given at KITP, April 7, 2015 and May 27, 2015}.

\bibitem{Maldacena:2016hyu}
Juan Maldacena and Douglas Stanford.
\newblock {Remarks on the Sachdev-Ye-Kitaev model}.
\newblock {\em Phys. Rev. D}, Vol.~94, No.~10, p. 106002, 2016.

\bibitem{Maldacena:2015waa}
Juan Maldacena, Stephen~H. Shenker, and Douglas Stanford.
\newblock {A bound on chaos}.
\newblock {\em JHEP}, Vol.~08, p. 106, 2016.

\bibitem{Cotler:2016fpe}
Jordan~S. Cotler, Guy Gur-Ari, Masanori Hanada, Joseph Polchinski, Phil Saad,
  Stephen~H. Shenker, Douglas Stanford, Alexandre Streicher, and Masaki Tezuka.
\newblock {Black Holes and Random Matrices}.
\newblock {\em JHEP}, Vol.~05, p. 118, 2017.
\newblock [Erratum: JHEP 09, 002 (2018)].

\bibitem{Berkooz:2018jqr}
Micha Berkooz, Mikhail Isachenkov, Vladimir Narovlansky, and Genis Torrents.
\newblock {Towards a full solution of the large N double-scaled SYK model}.
\newblock {\em JHEP}, Vol.~03, p. 079, 2019.

\bibitem{Berkooz:2018qkz}
Micha Berkooz, Prithvi Narayan, and Joan Simon.
\newblock {Chord diagrams, exact correlators in spin glasses and black hole
  bulk reconstruction}.
\newblock {\em JHEP}, Vol.~08, p. 192, 2018.

\bibitem{Okuyama:2022szh}
Kazumi Okuyama.
\newblock {Hartle-Hawking wavefunction in double scaled SYK}.
\newblock {\em JHEP}, Vol.~03, p. 152, 2023.

\bibitem{Okuyama:2023bch}
Kazumi Okuyama and Kenta Suzuki.
\newblock {Correlators of double scaled SYK at one-loop}.
\newblock {\em JHEP}, Vol.~05, p. 117, 2023.

\bibitem{Okuyama:2023iwu}
Kazumi Okuyama.
\newblock {High temperature expansion of double scaled SYK}.
\newblock {\em Phys. Lett. B}, Vol. 843, p. 138036, 2023.

\bibitem{Okuyama:2023byh}
Kazumi Okuyama.
\newblock {End of the world brane in double scaled SYK}.
\newblock {\em JHEP}, Vol.~08, p. 053, 2023.

\bibitem{Okuyama:2023kdo}
Kazumi Okuyama.
\newblock {Discrete analogue of the Weil-Petersson volume in double scaled
  SYK}.
\newblock {\em JHEP}, Vol.~09, p. 133, 2023.

\bibitem{Okuyama:2023aup}
Kazumi Okuyama and Takao Suyama.
\newblock {Solvable limit of ETH matrix model for double-scaled SYK}.
\newblock 11 2023.

\bibitem{Okuyama:2023yat}
Kazumi Okuyama.
\newblock {Matter correlators through wormhole in double-scaled SYK}.
\newblock 12 2023.

\bibitem{Mukhametzhanov:2023tcg}
Baur Mukhametzhanov.
\newblock {Large p SYK from chord diagrams}.
\newblock {\em JHEP}, Vol.~09, p. 154, 2023.

\bibitem{Parker:2018yvk}
Daniel~E. Parker, Xiangyu Cao, Alexander Avdoshkin, Thomas Scaffidi, and Ehud
  Altman.
\newblock {A Universal Operator Growth Hypothesis}.
\newblock {\em Phys. Rev. X}, Vol.~9, No.~4, p. 041017, 2019.

\bibitem{Balasubramanian:2022tpr}
Vijay Balasubramanian, Pawel Caputa, Javier~M. Magan, and Qingyue Wu.
\newblock {Quantum chaos and the complexity of spread of states}.
\newblock {\em Phys. Rev. D}, Vol. 106, No.~4, p. 046007, 2022.

\bibitem{Bhattacharjee:2022ave}
Budhaditya Bhattacharjee, Pratik Nandy, and Tanay Pathak.
\newblock {Krylov complexity in large q and double-scaled SYK model}.
\newblock {\em JHEP}, Vol.~08, p. 099, 2023.

\bibitem{Rabinovici:2023yex}
E.~Rabinovici, A.~S\'anchez-Garrido, R.~Shir, and J.~Sonner.
\newblock {A bulk manifestation of Krylov complexity}.
\newblock {\em JHEP}, Vol.~08, p. 213, 2023.

\bibitem{Susskind:2021esx}
Leonard Susskind.
\newblock {Entanglement and Chaos in De Sitter Space Holography: An SYK
  Example}.
\newblock {\em JHAP}, Vol.~1, No.~1, pp. 1--22, 2021.

\bibitem{Susskind:2022dfz}
Leonard Susskind.
\newblock {Scrambling in Double-Scaled SYK and De Sitter Space}.
\newblock 4 2022.

\bibitem{Susskind:2022bia}
Leonard Susskind.
\newblock {De Sitter Space, Double-Scaled SYK, and the Separation of Scales in
  the Semiclassical Limit}.
\newblock 9 2022.

\bibitem{Goel:2023svz}
Akash Goel, Vladimir Narovlansky, and Herman Verlinde.
\newblock {Semiclassical geometry in double-scaled SYK}.
\newblock {\em JHEP}, Vol.~11, p. 093, 2023.

\bibitem{Narovlansky:2023lfz}
Vladimir Narovlansky and Herman Verlinde.
\newblock {Double-scaled SYK and de Sitter Holography}.
\newblock 10 2023.

\bibitem{Rahman:2023pgt}
Adel~A. Rahman and Leonard Susskind.
\newblock {Comments on a Paper by Narovlansky and Verlinde}.
\newblock 12 2023.

\bibitem{Sachdev:2015efa}
Subir Sachdev.
\newblock {Bekenstein-Hawking Entropy and Strange Metals}.
\newblock {\em Phys. Rev. X}, Vol.~5, No.~4, p. 041025, 2015.

\bibitem{Gu:2019jub}
Yingfei Gu, Alexei Kitaev, Subir Sachdev, and Grigory Tarnopolsky.
\newblock {Notes on the complex Sachdev-Ye-Kitaev model}.
\newblock {\em JHEP}, Vol.~02, p. 157, 2020.

\bibitem{Berkooz:2020uly}
Micha Berkooz, Vladimir Narovlansky, and Himanshu Raj.
\newblock {Complex Sachdev-Ye-Kitaev model in the double scaling limit}.
\newblock {\em JHEP}, Vol.~02, p. 113, 2021.

\bibitem{Fu:2016vas}
Wenbo Fu, Davide Gaiotto, Juan Maldacena, and Subir Sachdev.
\newblock {Supersymmetric Sachdev-Ye-Kitaev models}.
\newblock {\em Phys. Rev. D}, Vol.~95, No.~2, p. 026009, 2017.
\newblock [Addendum: Phys.Rev.D 95, 069904 (2017)].

\bibitem{Li:2017hdt}
Tianlin Li, Junyu Liu, Yuan Xin, and Yehao Zhou.
\newblock {Supersymmetric SYK model and random matrix theory}.
\newblock {\em JHEP}, Vol.~06, p. 111, 2017.

\bibitem{Berkooz:2020xne}
Micha Berkooz, Nadav Brukner, Vladimir Narovlansky, and Amir Raz.
\newblock {The double scaled limit of Super--Symmetric SYK models}.
\newblock {\em JHEP}, Vol.~12, p. 110, 2020.

\bibitem{Biggs:2023mfn}
Anna Biggs, Juan Maldacena, and Vladimir Narovlansky.
\newblock {A supersymmetric SYK model with a curious low energy behavior}.
\newblock 9 2023.

\bibitem{Gu:2016oyy}
Yingfei Gu, Xiao-Liang Qi, and Douglas Stanford.
\newblock {Local criticality, diffusion and chaos in generalized
  Sachdev-Ye-Kitaev models}.
\newblock {\em JHEP}, Vol.~05, p. 125, 2017.

\bibitem{Berkooz:2016cvq}
Micha Berkooz, Prithvi Narayan, Moshe Rozali, and Joan Sim\'on.
\newblock {Higher Dimensional Generalizations of the SYK Model}.
\newblock {\em JHEP}, Vol.~01, p. 138, 2017.

\bibitem{Jian:2017unn}
Shao-Kai Jian and Hong Yao.
\newblock {Solvable Sachdev-Ye-Kitaev models in higher dimensions: from
  diffusion to many-body localization}.
\newblock {\em Phys. Rev. Lett.}, Vol. 119, No.~20, p. 206602, 2017.

\bibitem{Jian:2017jfl}
Chao-Ming Jian, Zhen Bi, and Cenke Xu.
\newblock {A model for continuous thermal Metal to Insulator Transition}.
\newblock {\em Phys. Rev. B}, Vol.~96, No.~11, p. 115122, 2017.

\bibitem{Song:2017pfw}
Xue-Yang Song, Chao-Ming Jian, and Leon Balents.
\newblock {Strongly Correlated Metal Built from Sachdev-Ye-Kitaev Models}.
\newblock {\em Phys. Rev. Lett.}, Vol. 119, No.~21, p. 216601, 2017.

\bibitem{Haldar:2017pyx}
Arijit Haldar, Sumilan Banerjee, and Vijay~B. Shenoy.
\newblock {Higher-dimensional Sachdev-Ye-Kitaev non-Fermi liquids at Lifshitz
  transitions}.
\newblock {\em Phys. Rev. B}, Vol.~97, No.~24, p. 241106, 2018.

\bibitem{Xu:2020shn}
Shenglong Xu, Leonard Susskind, Yuan Su, and Brian Swingle.
\newblock {A Sparse Model of Quantum Holography}.
\newblock 8 2020.

\bibitem{Garcia-Garcia:2020cdo}
Antonio~M. Garc\'\i{}a-Garc\'\i{}a, Yiyang Jia, Dario Rosa, and Jacobus J.~M.
  Verbaarschot.
\newblock {Sparse Sachdev-Ye-Kitaev model, quantum chaos and gravity duals}.
\newblock {\em Phys. Rev. D}, Vol. 103, No.~10, p. 106002, 2021.

\bibitem{Tezuka:2022mrr}
Masaki Tezuka, Onur Oktay, Enrico Rinaldi, Masanori Hanada, and Franco Nori.
\newblock {Binary-coupling sparse Sachdev-Ye-Kitaev model: An improved model of
  quantum chaos and holography}.
\newblock {\em Phys. Rev. B}, Vol. 107, No.~8, p. L081103, 2023.

\bibitem{Anegawa:2023vxq}
Takanori Anegawa, Norihiro Iizuka, Arkaprava Mukherjee, Sunil~Kumar Sake, and
  Sandip~P. Trivedi.
\newblock {Sparse random matrices and Gaussian ensembles with varying
  randomness}.
\newblock {\em JHEP}, Vol.~11, p. 234, 2023.

\bibitem{Anegawa:2023tgp}
Takanori Anegawa, Norihiro Iizuka, and Sunil~Kumar Sake.
\newblock {The local SYK model and its triple-scaling limit}.
\newblock {\em JHEP}, Vol.~10, p. 160, 2023.

\bibitem{erdHos2014phase}
L{\'a}szl{\'o} Erd{\H{o}}s and Dominik Schr{\"o}der.
\newblock Phase transition in the density of states of quantum spin glasses.
\newblock {\em Mathematical Physics, Analysis and Geometry}, Vol.~17, No. 3-4,
  pp. 441--464, 2014.

\bibitem{Baldwin:2019dki}
C.~L. Baldwin and B.~Swingle.
\newblock {Quenched vs Annealed: Glassiness from SK to SYK}.
\newblock {\em Phys. Rev. X}, Vol.~10, No.~3, p. 031026, 2020.

\bibitem{Swingle:2023nvv}
Brian Swingle and Michael Winer.
\newblock {A Bosonic Model of Quantum Holography}.
\newblock 11 2023.

\bibitem{spencerasymptopia}
J.H. Spencer and L.~Florescu.
\newblock {\em Asymptopia}.
\newblock Student mathematical library. American Mathematical Society.

\end{thebibliography}
\bibliographystyle{junsrt}

\end{document}